# A criterion for slip transfer at grain boundaries in Al

R. Alizadeh[1,2], M. Peña-Ortega[3], T. R. Bieler[4], J. LLorca[1,3,*]

[1]IMDEA Materials Institute, C/Eric Kandel 2, 28906 – Getafe, Madrid, Spain.
[2]Materials Science & Engineering Department, Sharif University of Technology, Tehran, Iran.
[3]Department of Materials Science, Polytechnic University of Madrid. E. T. S. de Ingenieros de Caminos, 28040 Madrid, Spain.
[4]Department of Chemical Engineering and Materials Science, Michigan State University, East Lansing, MI 48824-1226, USA.

*Corresponding author: javier.llorca@imdea.org

**Abstract**

The slip transfer phenomenon was studied at the grain boundaries of pure Aluminum by means of slip trace analysis. Either slip transfer or blocked slip was analyzed in more than 250 grain boundaries and the likelihood of slip transfer between two slip systems across the boundary was assessed. The experimental results indicate that slip transfer was very likely to occur if the residual Burgers vector, $\Delta b$, was below $0.35b$ and the Luster- Morris parameter was higher than 0.9, and that the ratio of the Luster-Morris parameter and the residual Burgers vector has a threshold above which slip transfer is probable.

**Keywords**: Grain boundaries; Slip; Aluminum; Dislocation

The mechanical properties of polycrystals are controlled to a large extent by the boundaries between grains with different crystal orientations. Grain boundaries (GBs) usually lead to the formation of dislocation pile-ups, but sometimes they can absorb and re-emit dislocations, or dislocations pass through the boundary, as reviewed in [1]. These processes have a large influence on the yield strength, the nucleation of twins, the generation of cracks during fatigue and creep, etc., but reliable criteria to predict the dislocation/GB interactions are missing due to the complexity and variability of the GB structure. This information is particularly critical to develop sound microstructure-based models of polycrystal deformation. In particular, crystal plasticity has demonstrated its capability to accurately reproduce the kinematic driving forces for slip in polycrystals [2,3] but standard crystal plasticity simulations assume that the GBs have no resistance to the transfer of the strain. This limitation may be overcome with the introduction of a length scale in the simulations through strain gradient plasticity models that can account for the development of dislocation pile-ups (in the form of geometrically necessary dislocations) due to the incompatibility of the deformation between neighboring grains [4–7]. Nevertheless, this approach does not take into account the geometry of slip and of the GB, which are known to play a critical role in slip transfer. More recently, the effect of grain boundaries on the response of polycrystals was simulated using a dislocation-based crystal plasticity model in which the rate of dislocation storage is not constant within the grain but increases as the distance to the GB decreases, simulating the formation of pile-ups [8-9]. This approach provided good predictions of the effect of grain size on the strength of FCC polycrystals but tended to overestimate the strengthening for small grain sizes (< 20 µm) because it was assumed that slip transfer was blocked in all GBs. Further refinement of the model [10] has shown that it is possible to allow or impede slip transfer across the boundary based upon the geometry of the slip systems across the boundaries, leading to better agreement with the experimental data. Slip transfer has been introduced into other models, e.g. [11-13]. Nevertheless, a reliable criterion that uses the misorientation between slip planes and Burgers vectors across the boundary to predict the likelihood of slip transfer is not available and this is the main objective of this investigation.

To this end, slip transfer across GBs was studied in high purity (99.9995%) polycrystalline Al foils of 200 µm in thickness purchased from Alfa Aesar. They were annealed at 360 ºC for 60 minutes to reach an average grain size of 390 ± 30 µm. Dog-bone tensile specimens 1 mm wide were extracted using electro-discharge machining with the tensile axis at 0º and 45° directions from the rolling direction. The grain orientation on both surfaces of the specimen was determined using electron back-scatter diffraction (EBSD). Approximately half of the grains on the front surface are also present on the back surface. More details about the sample preparation can be found in our previous publication on cube-oriented samples [11]. Tensile direction crystal orientation maps from the central region of two samples are illustrated in Fig. 1. Most of the grains in Fig. 1a are red-orange, indicating a near-cube orientation, while some grains with different colors (orientations) are evident. The 45° sample illustrated in Fig. 1b, has a rotated cube orientation, showing a prevalent green or blue color.

The samples were deformed in tension at room temperature using a micro-tensile testing machine (Kammrath and Weiss) up to an applied strain of 4% at a strain rate of ≈ $10^{-3}$ s$^{-1}$. After the tests, slip traces were evident in nearly every grain of the surface of the foils which were examined using secondary electron imaging. Slip transfer was assessed in more than 250 GBs which were classified in two groups: GBs where slip transfer was not observed and GBs where slip transfer was convincingly observed. The latter was indicated when imposing and receiving slip traces in grains were clearly correlated, and there was little topography along the grain boundary (which indicates relatively homogeneous deformation in both grains). Examples of these two cases are shown in Fig. 2a. More details about the method used to analyze slip traces and several examples can be found in our previous publication [14]. Moreover, the slip systems with the highest Schmid factor (SF) were identified for each grain across the boundary from the EBSD information assuming that all the grains were subjected to uniaxial tension along the loading axis. Out of the 12 slip systems in each grain, there are only four slip planes. Near-cube orientations (Fig. 1a) have 8 slip systems with high Schmid factors, 4 of which have Burgers vectors that are nearly parallel to the surface, making evidence for their operation uncertain. In contrast, rotated near-cube orientations (Fig. 1b) have only four high Schmid factor slip systems (all of which have large out-of-plane components), and eight systems with very low Schmid factors. The slip systems that are most probably active and whose slip traces should be visible, and the correspondence between the observed and computed slip systems enables identification of the slip system(s) that dominate the deformation within each grain.

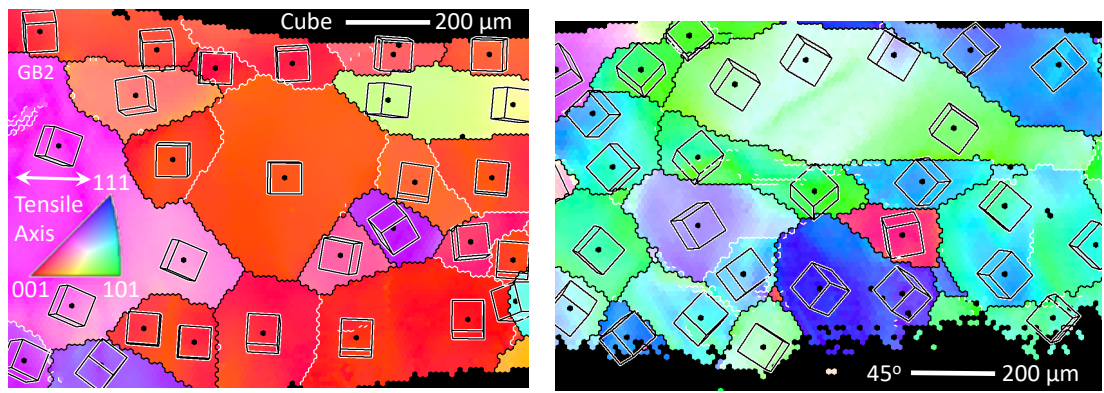

Fig. 1. Crystal orientation maps obtained by EBSD of the front surface of the samples. (a) Near-cube sample. (b) Rotated-cube (45°) sample. The EBSD orientation maps are colored based upon the horizontal (Y) tensile axis inverse pole figure. The crystal orientations are also indicated with unit cell prisms. High angle boundaries (>15°) are marked with black lines, and low angle boundaries between 6º and 15° are marked with white lines.

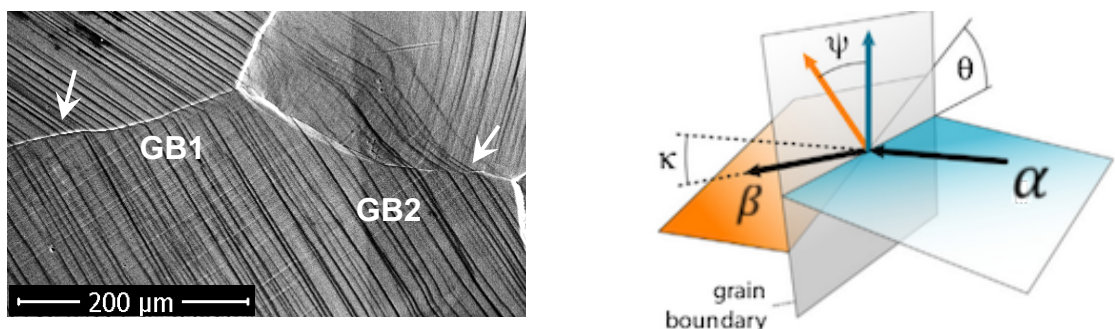

Fig. 2. (a) GBs showing different slip transfer character at arrows: blocked slip (GB1), slip transfer (GB2, marked in upper left corner of Figure 1). (b) Angles and vectors used to evaluate the likelihood of slip transfer across a GB from slip system α to β [10].

Different metrics can be used to assess the likelihood of slip transfer at the GB. The LRB parameter (based upon the angle between two Burgers vector directions $\kappa$ and the intersection angle $\theta$ between the two slip planes in the boundary plane, Fig. 2b) can account for slip transmission when the GB inclination is known, typically in transmission electron microscopy investigations [15]. When the GB inclination is not known from a surface EBSD map, for example, the Luster-Morris parameter $m'_{\alpha\beta}$ [1,16], based upon the angles between the slip plane normal directions $\psi$ and the Burgers vector directions $\kappa$ (Fig. 2b), also provides a good predictor of slip transfer in Al [10] and Ti [17]. It is expressed as

$$m'_{\alpha\beta} = \cos\psi \cos\kappa \qquad (1)$$

for each combination of slip systems $\alpha$ and $\beta$ (Fig. 2b). Slip transfer will be unlikely if the driving force for slip across the boundary is very low in either grain (low Schmid factor). Thus, another meaningful metric for slip transfer is given by

$$m'_{\alpha\beta}(SF_\alpha + SF_\beta) > f(\theta, microstructure) \qquad (2)$$

where $SF_\alpha$ and $SF_\beta$ are the Schmid factors in slip systems $\alpha$ and $\beta$ and $f(\theta)$ represents a function defining a threshold that depends on the misorientation angle $\theta$ between the two grains, grain orientations and the microstructure [10].

In addition to the Luster-Morris parameter, slip transfer is more likely when the residual Burgers vector left at the boundary is low [1,18]. The magnitude of the residual Burgers vector after slip transfer, $\Delta b$, can be calculated as

$$\Delta b = |\boldsymbol{b}_\alpha - \boldsymbol{b}_\beta| = 2\sin\left(\frac{\kappa}{2}\right)b \quad \text{where} \quad \kappa \leq \pi/2 \qquad (3)$$

The ability of these three criteria to predict slip transfer was compared with experimental evidence of slip transfer and blocked slip across GBs in Al. In the cases where slip transfer was observed, the values for the different criteria were computed from the slip systems where slip transfer was observed, which usually had high SFs. In the cases where slip transfer did not occur, the highest values of $m'_{\alpha\beta}$ and the minimum residual Burgers vector among highly favored slip systems were considered to compute the criteria for predicting slip transfer.

The $m'_{\alpha\beta}$ values for observed instances of slip transfer and the maximum $m'_{\alpha\beta}$ values among favored slip systems with SF > 0.25 for blocked slip are plotted against misorientation angle in Figs. 3a and 3b for the boundaries where slip transfer was observed and blocked, respectively (about 27% of the boundaries in the cube orientation and 8% of the boundaries in the rotated cube samples show some features of slip transfer, but are not convincing, so these data are omitted). Results from the near-cube and rotated-cube specimens are shown in both figures using appropriate symbols. There are subtle differences between the cube oriented and 45° rotated oriented samples that can be discerned by close inspection of the symbols used, but they are plotted in the same color so that the behavior of the population as a whole is evident. Observed slip transfer *usually* occurred with the highest $m'_{\alpha\beta}$ value, or on a pair of slip systems with $m'_{\alpha\beta}$ values close to the highest value (i.e., near the trend with downward curvature apparent in Figs. 3a and 3b up to about 35°). Slip transfer is favored for low angle boundaries, which naturally have multiple high $m'_{\alpha\beta}$ values. There are a few instances where no slip transfer was observed for low angle boundaries with high $m'$ values, which may be a consequence of larger deviations from the cube orientation or deviations from a uniaxial stress state in some regions of the microstructure [19]. Slip transfer clearly occurred for some boundaries with high misorientation angles, but it was rare. The scatter for blocked slip observations (Fig. 3b) became large with misorientations > 30°.

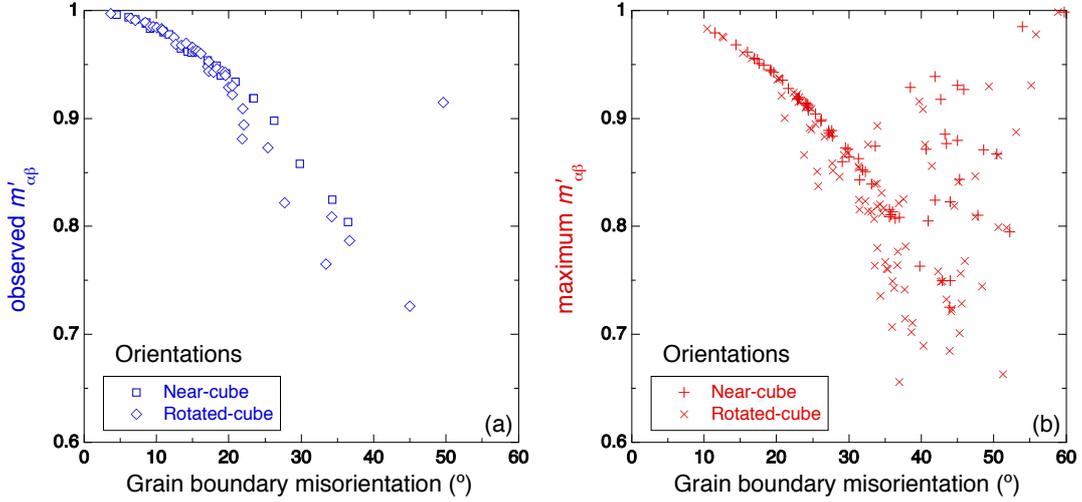

Fig. 3. Metrics for slip transfer as a function of the GB misorientation angle: (a) Observed $m'_{\alpha\beta}$ in the GBs where slip transfer was observed. (b) Maximum $m'_{\alpha\beta}$ among favored slip systems in GBs that blocked slip.

A metric that combines two factors necessary for slip transfer results from multiplying the sum of the SFs in systems α and β by the corresponding $m'_{\alpha\beta}$. This is plotted in Fig. 4a as a function of the GB misorientation for the samples from both the near-cube and rotated-cube populations using the maximum $m'_{\alpha\beta}$ for blocked slip boundaries for which $SF_\alpha$ and $SF_\beta > 0.25$. The black lines in Fig. 4a identify empirical thresholds above which slip transfer is observed at a high percentage (values in blue) and the corresponding percentage for blocked slip transfer below the line (values in red). The threshold for the near-cube data indicates that 68% of the observed slip transfer observations are above the threshold while 99% of the blocked slip boundaries are below. A lower threshold would raise the fraction of observed slip transfer events, but would lower the fraction of blocked slip observations. The threshold for the rotated-cube samples is lower and has a steeper slope, consistent with the fact that there are fewer slip systems with high SFs.

The influence of parameters $m'_{\alpha\beta}$ and $\Delta b$, is assessed using the $m'_{\alpha\beta}/\Delta b$ ratio in Fig. 4b, as slip transfer should be facilitated with high $m'_{\alpha\beta}$ and low $\Delta b$. The statistics show similar separation of the slip transfer and blocked slip populations for the near-cube data, and better separation for the rotated cube data than Fig. 4a. As the ratio does not depend on knowing the local stress state or deformation history, the uncertainty associated with Schmid factors based upon the global stress tensor is eliminated.

The assessment metrics in Figs. 4a and b is intrinsically different for observed slip transfer and blocked slip, as instances of observed slip transfer can be correlated with observed slip systems, whereas only the potential for slip transfer can be assessed when no slip transfer occurred. To examine this distinction more carefully, Fig. 4c compares the relationship between $m'_{\alpha\beta}$ and $\Delta b$ for both populations in a more similar manner, where the maximum value of $m'_{\alpha\beta}$ for blocked slip is limited to the *observed* active slip systems in both grains. These maximum $m'_{\alpha\beta}$ values are different from those in Fig 4a,b (where the maximum $m'_{\alpha\beta}$ among *all* pairs of favored slip systems across the GB were considered for blocked slip boundaries). Fig. 4c shows a much better separation of slip transfer and blocked slip using $m'_{\alpha\beta} > 0.88$ and $\Delta b/b < 0.38$ (which best separates the two populations), and this provides better separation than the ratio plot in Fig. 4b for the near-cube data, and a possible improvement for the rotated cube data (plotting the maximum $m'_{\alpha\beta}$ slip blocked data in the same manner leads to the poorest statistical separation and is not shown). Plotting the data of Fig. 4c in the same way as Fig. 4b results in Fig. 4d provides the best statistical separation of the two populations, where 95% of the slip transfer and blocked slip data are on either side of the empirical threshold line for the near-cube data, and similarly improved statistics for the rotated-cube data.

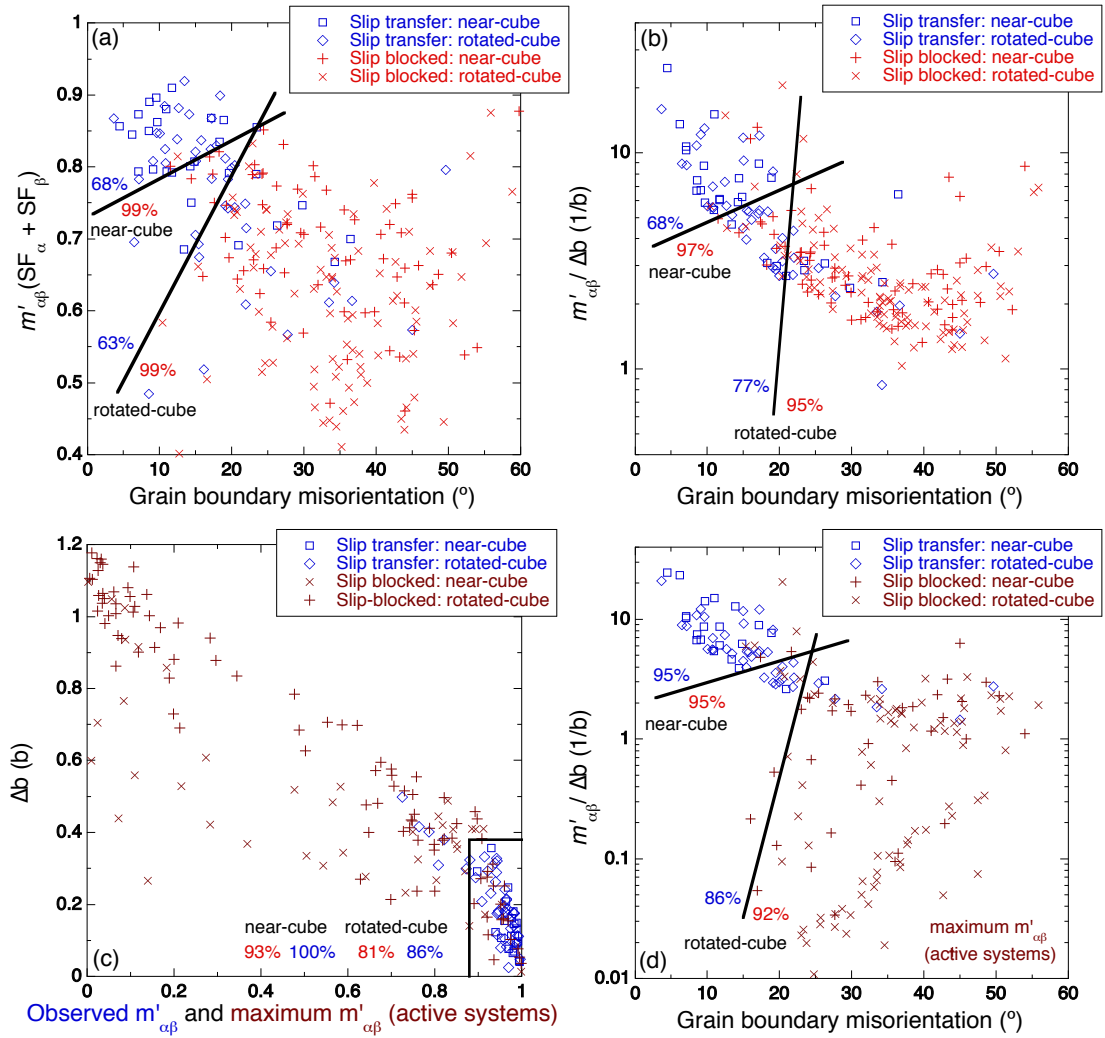

Fig. 4. Metrics of slip transfer as a function of various geometrical criteria. (a) Observed $m'_{\alpha\beta}(SF_\alpha+SF_\beta)$ for slip transfer and maximum $m'_{\alpha\beta}(SF_\alpha+SF_\beta)$ for blocked slip *vs.* GB misorientation. (b) $m'_{\alpha\beta}/\Delta b$ (where $\Delta b$ is expressed as a fraction of $b$) *vs.* GB misorientation. (c) $\Delta b$ (expressed as a fraction of $b$) *vs.* observed $m'_{\alpha\beta}$ (slip transfer) or maximum observed $m'_{\alpha\beta}$ among the active slip systems (slip blocking). (d) *Idem* as (b) but considering only the maximum observed $m'_{\alpha\beta}$ among the active slip systems for slip blocking. Thick black lines identify empirical thresholds between predominant slip transfer and blocked slip for cube-oriented and rotated-cube populations; percents of slip transfer are blue numbers, and percents of blocked slip are red numbers.

The thresholds indicated in Fig. 4 can be easily implemented into crystal plasticity codes, as they are strictly geometric (the SF sum could be computed based upon *local* stress states or a similar metric based upon the amounts of local accumulate shear on slip systems could be used). Furthermore, crystal plasticity models provide evolution of local crystal orientations, so the effect of grain boundary slip transparency could be examined with simulations where the grain boundary properties evolve by assessing $m'_{\alpha\beta}$ and $\Delta b$ as the material deforms.

While the $m'_{\alpha\beta}/\Delta b$ threshold appears to be most useful, there are two open questions that require further investigations. First, the origin of outliers in the two populations needs to be examined. As it is very difficult to obtain localized stress and strain information in experiments, knowing how the local stress and local strain evolve using simulations of these experiments could provide important insights that could enable a rational explanation for these outlier data. Second, it is not yet clear how the microstructure and grain orientations affect the threshold function. While this is clearly related to the number of potentially active slip systems, the experimental observations show there are more potentially activated slip systems than activated slip systems. This suggests that the details of microyielding may have a significant impact on activation of subsequent slip, i.e., the first activated slip system may make later activation of other systems, even if they are more highly stressed, difficult, due to latent hardening effects [20].

In summary, experimental observations of slip transfer in Al polycrystals have shown that slip transfer tends to occur predominantly at lower angle grain boundaries. The analysis of > 250 grain boundaries indicates that slip transfer is likely to occur when $m'_{\alpha\beta} > 0.9$ and $\Delta b < 0.35b$, or at values of $m'_{\alpha\beta}/\Delta b$ that exceed a threshold. Such geometrical criteria can be implemented into crystal plasticity models to account for the effect of grain boundaries on the deformation of polycrystals [8-10]. It is expected that this strategy will provide a better description of the evolving heterogeneous deformation of crystals near grain boundaries as well as the associated local hardening and resulting stress concentrations within polycrystals resulting from GBs.

**Acknowledgments**

This investigation was supported by the European Research Council (ERC) under the European Union's Horizon 2020 research and innovation programme (Advanced Grant VIRMETAL, grant agreement No. 669141. Additional support from the HexaGB project supported by the Spanish Ministry of Science (reference RTI2018-098245) is also acknowledged. TRB also acknowledges support from the US Department of Energy Office of Basic Science via grant DE-FG02-09ER46637.